\newcommand{\bluef}[1]{{\textcolor{black}{Figure~#1}}}
\newcommand{\bluet}[1]{{\textcolor{black}{Table~#1}}}
\newcommand{\invcm}{{cm$^{-1}$}}

\documentclass[showkeys,showpacs,prb,twocolumn, superscriptaddress,bibnotes]{revtex4}
\usepackage{color}
\usepackage{graphicx}

\begin{document}

\title{Strain dependence of the heat transport properties of graphene nanoribbons}

\author{Pei Shan Emmeline Yeo}
\affiliation
{Department of Chemistry, National University of Singapore,
3 Science Drive 3, Singapore 117543}
\affiliation
{Institute of High Performance Computing, Agency
for Science, Technology and Research, 1 Fusionopolis Way, \#16-16
Connexis, Singapore 138632}
\author{Kian Ping Loh}
\affiliation
{Department of Chemistry, National University of Singapore,
3 Science Drive 3, Singapore 117543}
\author{Chee Kwan Gan}
\email{ganck@ihpc.a-star.edu.sg}
\affiliation
{Institute of High Performance Computing, Agency
for Science, Technology and Research, 1 Fusionopolis Way, \#16-16
Connexis, Singapore 138632}

\begin{abstract}
Using a combination of accurate density-functional theory and a nonequilibrium
Green function's method, we calculate the ballistic thermal conductance
characteristics of tensile-strained armchair (AGNR) and zigzag (ZGNR)
edge graphene nanoribbons, with widths between $3-50$~\AA{}. The
optimized lateral lattice constants for AGNRs of different widths
display a three-family behavior when the ribbons are grouped according
to $N$ modulo 3, where $N$ represents the number of carbon atoms
across the width of the ribbon. Two lowest-frequency out-of-plane
acoustic modes play a decisive role in increasing the thermal conductance
of AGNR-$N$ at low temperatures. At high temperatures the effect
of tensile strain is to reduce the thermal conductance of AGNR-$N$
and ZGNR-$N$. These results could be explained by the changes in
force constants in the in-plane and out-of-plane directions with the
application of strain. This fundamental atomistic understanding of
the heat transport in graphene nanoribbons paves a way to effect changes
in their thermal properties via strain at various temperatures.

\end{abstract}


\maketitle

\section{Introduction}

Recently there is a surge in research activities on heat transport
through nanostructures as evidenced by the emergence of a few review
papers.\citep{Dubi11v83,Wang08v62,Balandin11v10} The reasons for
this change abound. The first is related to heat management in nanoelectronic
circuits,\citep{Ghosh08v92} since the miniaturization of electronic
devices demands efficient dissipation of heat. The second is related
to the utilization of the thermoelectric effect\citep{Harman02v297}
to harness heat in nanostructures that may help in
alleviating the worldwide energy problem. Graphene and its derivatives
such as graphene nanoribbons (GNRs) are among the most promising materials
in these respects. Various experimental values for the heat conductivity
of graphene have been reported, e.g., $4840-5300$~Wm$^{-1}$K$^{-1}$
(Ref.~\citenum{Balandin08v8}), $600-630$~Wm$^{-1}$K$^{-1}$
(Ref.~\citenum{Faugeras2010}), and $1400-2500$~Wm$^{-1}$K$^{-1}$
(Ref.~\citenum{Caiw2010}). This points to the fact that high heat
conductivity is expected for graphene (in stark contrast to, e.g.,
the heat conductivity of Ag which is only $\sim430$~Wm\textsuperscript{-1}K\textsuperscript{-1}
at room temperature). The high thermal conductance of graphene has
made it very popular for use as a filler in thermal interface materials.\citep{Balandin11v10}
For example, the heat conductivity of epoxy resins was improved by
$30$~times upon addition of $25$~vol\% graphene additive,\citep{Yu2007}
and by $2.6$~times when $2$~wt\% of graphene was added to polystyrene.\citep{Fang2010}

Graphene could also be potentially used as a thermoelectric material
to generate thermoelectric power.\citep{Balandin11v10} The efficiency
of thermoelectric materials can be quantified using the thermoelectric
figure of merit ZT$=S^{2}G_{e}T/(\sigma_{el}+\sigma_{ph})$, where
$S$ is the Seebeck coefficient (also known as the thermopower), $G_{e}$
is the electronic conductance, $T$ is temperature and $\sigma_{el}$
($\sigma_{ph}$) is the electronic (thermal) conductance. Graphene
has a superior\citep{Bolotin2008} electronic conductance $G_{e}$,
and a large\citep{Dragoman2007} theoretical value of $S\sim30$~mV/K.
Even though the experimental\citep{Wei2009} values of $S$ for graphene
are more modest ($40-80$~$\mu$V/K) compared to that for the inorganic\citep{Harman02v297,Hsu2004}
thermoelectric materials ($150-850$~$\mu$V/K), graphene might still
qualify as a good thermoelectric material if its high value of $(\sigma_{el}+\sigma_{ph})$
could be suppressed. Although graphene is a semi-metal, its heat conduction
is dominated by $\sigma_{ph}$ and not by $\sigma_{el}$ due to the
strong sp\textsuperscript{2}-hybridization that efficiently transmits
heat through lattice vibrations.\citep{Klemens2000a} Various ways
have been proposed to increase the phonon scattering centers in graphene,
e.g., by increasing the disorder at graphene edges,\citep{Savin2010}
by introducing isotopes in graphene,\citep{Hu2010} and by creating
vacancy defects in graphene.\citep{Haskins11v5} GNRs with vacancy
defects was predicted to have a ZT of up to $0.25$.\citep{Gunst2011}

The experimental demonstrations of the excellent heat properties of
graphene and GNRs have stimulated many theoretical works.\citep{Ghosh09v11,Nika09v79,Lan09v79,Xie11v23,Guo09v95,Wei11v22,Hu09v9,Munoz10v10,Tan11v11}
It is known that applying strain to graphene induces changes to the
electronic structure,\citep{Sun08v129} resistance,\citep{Kim09v457}
Raman spectra,\citep{Huang09v106} and thermal conductivity\citep{Li10v81}.
For GNRs, different theoretical approaches have been used to study
the heat properties of both the unstrained and strained GNRs. Thus
far, molecular dynamics (MD) studies concluded that both compressive
and tensile strains are detrimental to the heat conductivity of graphene\citep{Li10v81}
or GNRs.\citep{Guo09v95,Wei11v22,Gunawardana12v85}
Wei \emph{et al.}\citep{Wei11v22} and Gunawardana \emph{et al.}\citep{Gunawardana12v85}
concluded that the conductivity
of armchair edge GNRs is more sensitive toward strain than their zigzag
edge counterparts. Guo \emph{et al.}\citep{Guo09v95} used a slightly
different approach than that used in Ref.~\citenum{Wei11v22},
which resulted in slightly different but essentially similar predictions. 
We note that the MD method is well-suited for investigating heat
conduction in the diffusive regime and at high temperatures. However,
in the ballistic regime and at low temperatures, intricate quantum
mechanical effects come into play.\citep{Tan11v11} Since the phonon
mean free path in graphene is $\sim775$ nm at room temperature,\citep{Ghosh08v92}
the heat conduction is ballistic for small-scale graphene nanodevices. The phonon 
mean free path is reduced to $\sim20$~nm in presence of edge disorders\citep{Sevincli10v81,Sevincli10v82}.
Zhai \emph{et al.}\citep{Zhai2011}
addressed the thermal conductance of GNRs using a ballistic nonequilibrium
Green's function (NEGF) approach.\citep{Wang06v74,Wang07v75,Mingo06v74}
They extracted the force constants of strained graphene via the elasticity
theory and applied that to study strained GNRs. They concluded that
thermal conductance is enhanced with tensile strain, with an enhancement
ratio of up to 17\% and 36\% for zigzag edge and armchair edge GNRs,
respectively.\citep{Zhai2011} However, we note that a large 19\%
strain applied in Ref.~\citenum{Zhai2011} might put the applicability
of the elasticity theory in the high strain regime to a severe test.

In this work, we investigate the thermal conductance characteristics
of strained GNRs by using a combination of (1) density-functional
theory (DFT) that accurately treats the atomic and electronic structures
of sub-nanometer width GNRs, and (2) the NEGF method that has been
extensively used to study the heat\citep{Wang06v74,Mingo06v74} and
electron\citep{Brandbyge02v65} transport through nanostructures.
Our results may shed light on how the heat conductivity of graphene-polymer
composites could be affected under loading, and the possibility of
using strain to tune the thermal conductance of GNRs to improve its ZT
value.

\section{Models and Methodology}

\begin{figure}
\begin{raggedleft} \includegraphics[clip,width=6.5cm]{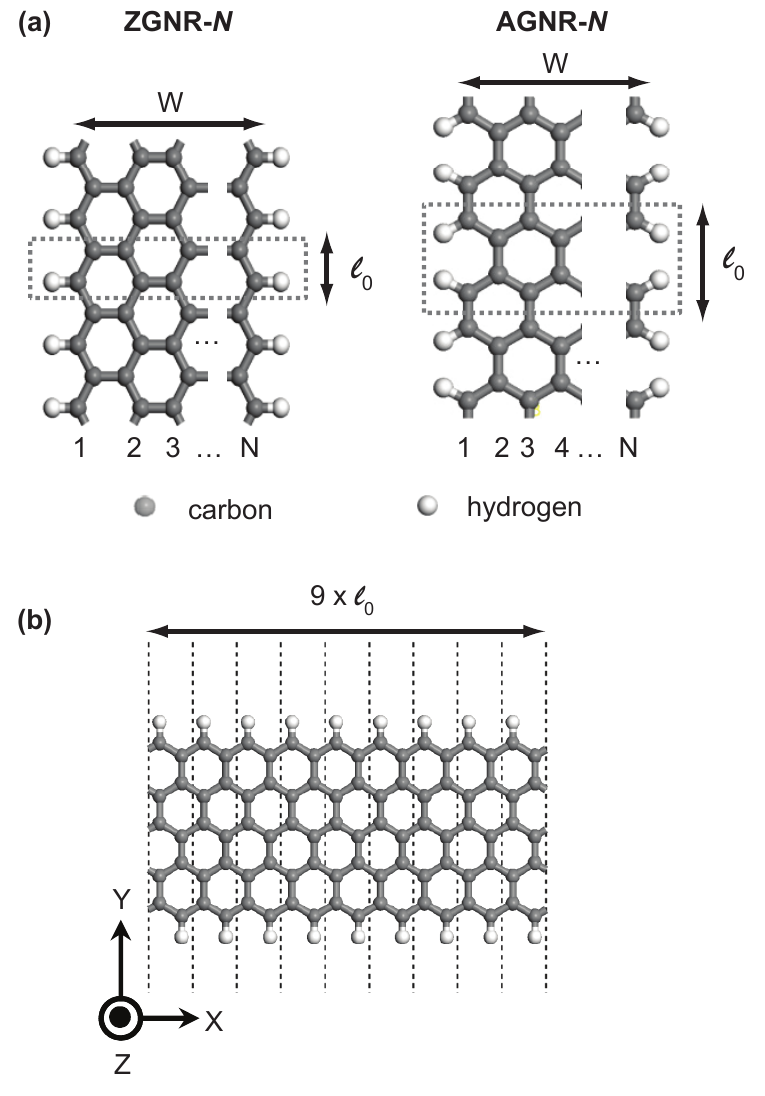} \caption{\textbf{(a)} The width $W$ of the zigzag (ZGNR-$N$) and armchair
(AGRNR-$N$) edge graphene nanoribbons is controlled by $N$ that
represents the number of carbon atoms across the width of the ribbon.
Hydrogen atoms are attached to the edge carbon atoms to terminate
the dangling bonds. For very large $W$, the optimized length $\ell_{0}$
of the primitive cell along the edge direction should approach $a'_{0}$
for ZGNR-$N$ and $a'_{0}\sqrt{3}$ for AGNR-$N$, where $a'_{0}$
is the lattice parameter of graphene. The primitive unit cells are
demarcated by dotted lines. \textbf{(c)} A supercell comprising of
nine primitive unit cells is constructed for the phonon calculations.
\label{fig:Naming_conv_and_simul_cell} }

\end{raggedleft} 
\end{figure}

In this work, we calculate the thermal conductance of the zigzag (ZGNR-$N$)
and armchair edge graphene nanoribbons (AGNR-$N$) as shown in \bluef{\ref{fig:Naming_conv_and_simul_cell}(a)}
using a combination of first-principles density-functional calculations
and the ballistic nonequilibrium Green's function (NEGF) method.\citep{Wang06v74,Wang07v75,Mingo06v74,Huang10v108}
The uniaxial strain imposed on the GNRs is described by the strain
parameter $\varepsilon=(\ell-\ell_{0})/\ell_{0}$, where $\ell$ ($\ell_{0}$)
is the length (relaxed length) of the ribbon along the edges. A tensile
(compressive) strain corresponds to $\varepsilon>0$ ($\varepsilon<0$).
We note that the GNR edges have compressive edge stresses\citep{Gan10v81}
that might cause the GNRs to buckle\citep{Bao09v4,Kumar10v82} that will lead to a decrease of
thermal conductance\citep{Li10v81,Guo09v95,Wei11v22} due to increased
phonon-phonon scattering. A proper treatment of buckled GNRs using
DFT involves many atoms in a supercell and this demands extensive
computing resources. Therefore we limit this work to studying the
effects of tensile strain on the flat GNRs. The thermal conductance
$\sigma(T,\varepsilon)$ at temperature $T$ and strain $\varepsilon$
is calculated from the Landauer expression, 
\begin{equation}
\sigma(T,\varepsilon)=\int_{0}^{\infty}d\nu\ h\nu\theta(\nu)\frac{\partial n_{B}(\nu,T)}{\partial T}
\end{equation}
or equivalently,\citep{Markussen08v8} 
\begin{equation}
\sigma(T,\varepsilon)=\frac{h^{2}}{kT^{2}}\int_{0}^{\infty}d\nu\ \nu^{2}\theta(\nu)\frac{e^{h\nu/kT}}{(e^{h\nu/kT}-1)^{2}}\label{eq:differentiated}
\end{equation}
where $n_{B}(\nu,T)=\frac{1}{e^{h\nu/kT}-1}$ is the Bose-Einstein
distribution for frequency $\nu$ and temperature $T$, $h$ ($k$)
is the Planck (Boltzmann) constant. The key quantity is the transmission
function $\theta(\nu)$ that may be calculated in general cases using
the nonequilbrium Green's function method\citep{Wang07v75} or by
counting the number of phonon bands at frequency $\nu$ for quasi-one-dimensional
periodic systems.\citep{Markussen08v8} We have used the latter approach
to get $\theta(\nu)$ due to its computational efficiency to treat
the problem at hand. It is interesting to note that at low temperatures
$T$, only the very low-frequency modes contribute to thermal conductance.
Therefore $\theta(\nu\rightarrow0)=N_{m}$ in eqn. \ref{eq:differentiated}
may be taken out of the integral sign and this leads to a quantization\citep{Rego98v81,Yamamoto04v92}
of the thermal conductance according to $\sigma(T,\varepsilon)=\frac{k^{2}T}{h}N_{m}\int_{0}^{\infty}du\frac{u^{2}e^{u}}{(e^{u}-1)^{2}}=N_{m}\frac{\pi^{2}k^{2}T}{3h}$.

We perform density-functional theory (DFT) calculations using the
SIESTA package.\citep{Soler2002} The local-density approximation
is used for the exchange-correlation functional. Double-$\zeta$ basis
sets and Troullier-Martins pseudopotentials are used for the C and
H atoms. We use a vacuum separation of 15~\AA{}\ in the $y$ and
$z$ directions consistent with the convention adopted in \bluef{\ref{fig:Naming_conv_and_simul_cell}(b)}.
The mesh cutoff is $400$~Ry. The atomic positions are relaxed using
the conjugate gradient algorithm with a force tolerance criterion
of $10^{-3}$~eV/\AA{}. As was demonstrated in Refs.~\citenum{Son06v97}
and ~\citenum{Gan10v81}, spin-polarization effects are particularly
important for ZGNRs. Therefore we perform spin-polarized (nonspin-polarized)
calculations for ZNGR-$N$ (AGNR-$N$).

Phonon dispersion relations of GNRs are calculated using the supercell
method.\citep{Kresse1995,Gan06v73,Zhao11v84} To minimize interactions
from the distant periodic images of a displaced atom from its equilibrium
position, a supercell of nine primitive cells sufficient for this
purpose is used.\citep{Tan11v11} We displace the $i$th atom in
a primitive cell from its equilibrium position by $\pm\delta_{i\alpha}=\pm0.015$~\AA{}
and evaluate the forces acting on the $j$th atom in the supercell
$F_{j\beta}(\pm\delta_{i\alpha})$ using the Hellmann-Feynman theorem.
$\alpha$ and $\beta$ denote the Cartesian directions. We then use
a finite central-difference scheme to evaluate the matrix elements
of the force constant matrix $K$, where $K_{i\alpha,\, j\beta}=\frac{\partial^{2}E}{\partial r_{i\alpha}\partial r_{j\beta}}=-\left[\frac{F_{j\beta}(+\delta_{i\alpha})-F_{j\beta}(-\delta_{i\alpha})}{2\delta_{i\alpha}}\right]$.
To reduce the number of static DFT calculations, we exploit the space
group operations of AGNR-$N$ and ZGNR-$N$ so that only atoms in
the inequivalent positions are displaced. The forces\citep{Kresse1995}
or interatomic force constants on the equivalent atoms are deduced
from that of the inequivalent atoms. 
AGNR-$N$ with $N=2(p+1)$ and $N=2p+1$ belong to the space group
number 51 and 47, respectively, where $p\ge1$ is a positive integer;
while ZGNR-$N$ with $N=2p$ and $N=2p+1$ belong to the space group
number 47 and 51, respectively. The unstrained (strained) graphene belongs to the space group number 191 (65).

\section{Results and Discussion}

\subsection{Optimized Lattice Parameter of GNRs}

\begin{figure}
\begin{raggedleft} \includegraphics[clip,width=7.2cm]{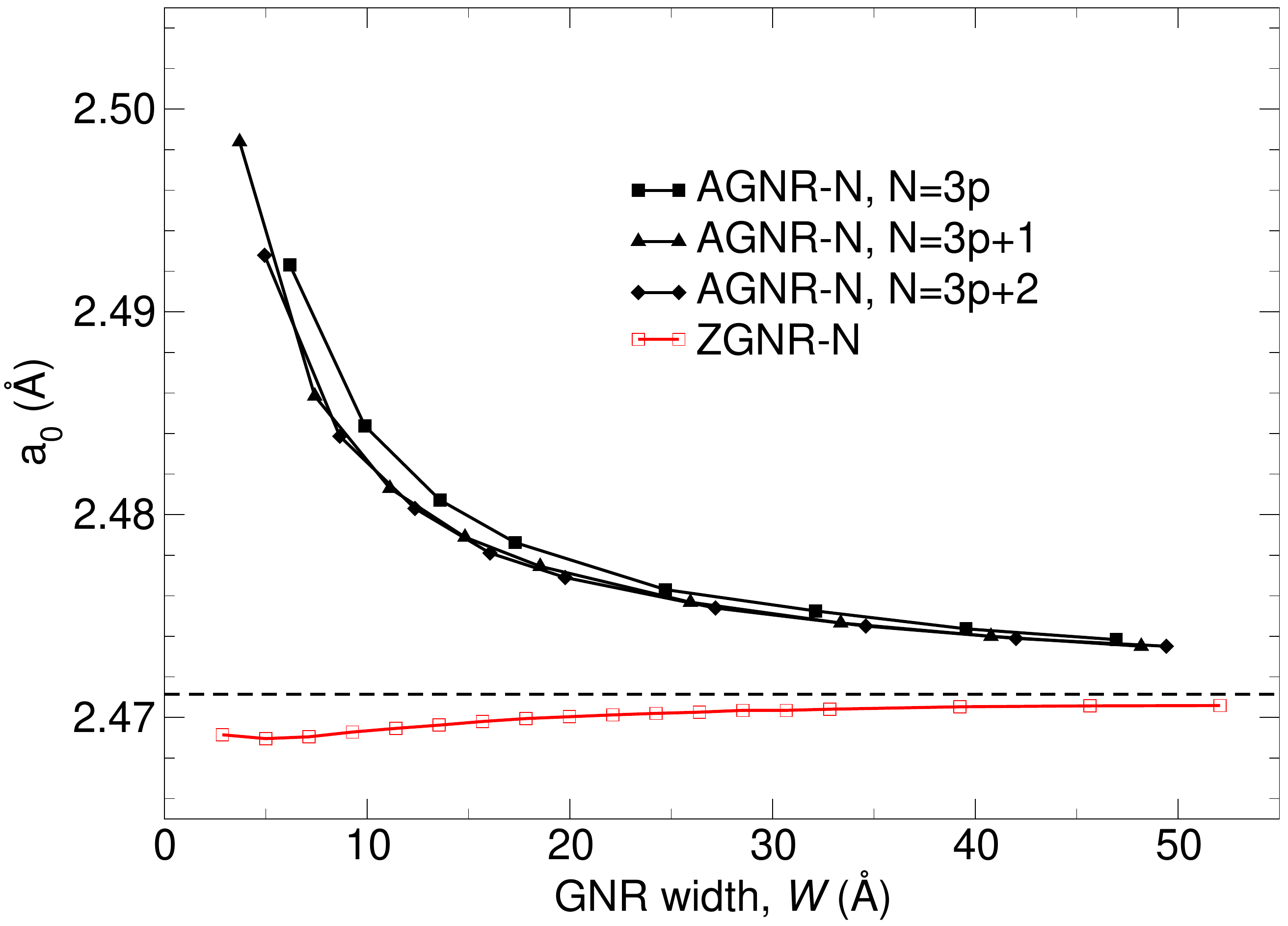}

\end{raggedleft}

\caption{The approach of the optimized lateral lattice parameter $a_{0}$ for
AGNR-$N$ ($N=4$ to $41$) and ZGNR-$N$ ($N=2$ to $25$) toward
$a'_{0}$, the optimized lattice parameter of graphene (denoted by
a horizontal dash line) as the width $W$ of the ribbon increases.
A three-family behavior is observed for AGNR-$N$. \label{fig:Lattice_param} }
\end{figure}

\begin{figure}
\begin{raggedleft} \includegraphics[clip,width=6.7cm]{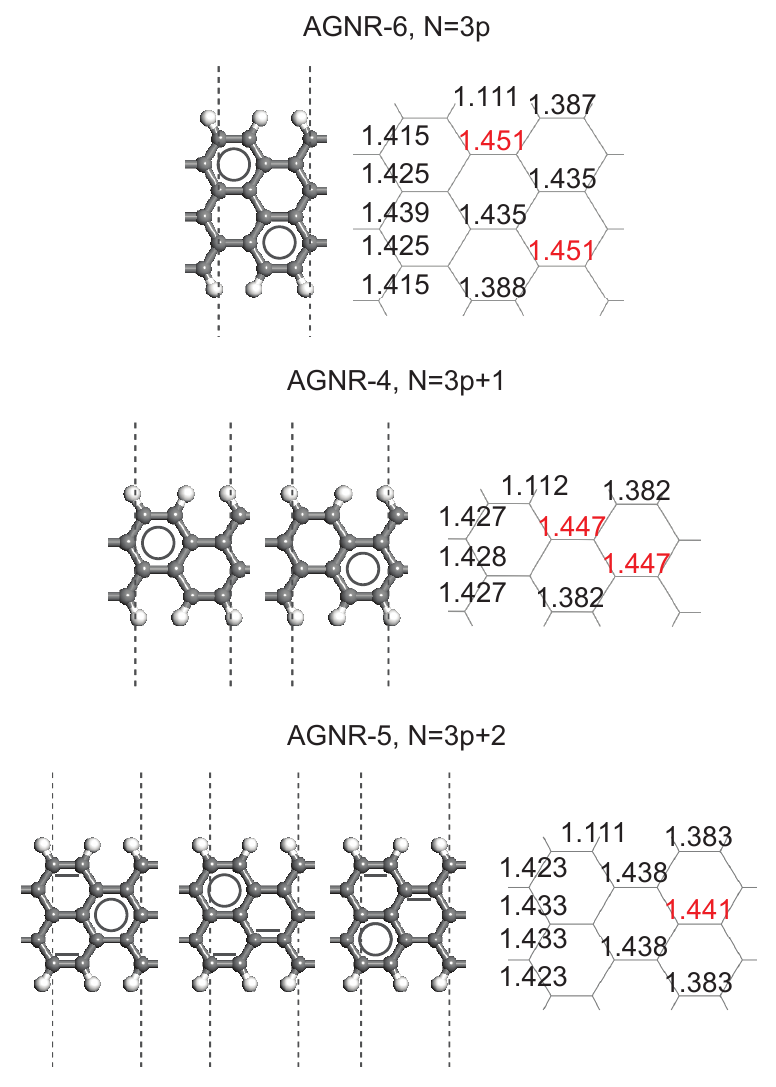}

\end{raggedleft}

\caption{The three families of AGNR-$N$: the $N=3p$ family has only one possible
Clar's structure, the $N=3p+1$ family has two Clar's structures and
the $N=3p+2$ family has more than two Clar's structures. The bond
lengths for each optimized AGNR are also shown. The longest C--C
bonds in each structure are highlighted in red. \label{fig:clar_struc} }
\end{figure}

Since this work concerns the effect of strain $\varepsilon=(\ell-\ell_{0})/\ell_{0}$
on the thermal conductance, we first need to obtain the optimized
length $\ell_{0}$ (see \bluef{\ref{fig:Naming_conv_and_simul_cell}})
of GNR-$N$ for different $N$ (or equivalently, width $W$). We obtain
$\ell_{0}$ for each $N$ by performing atomic relaxation of GNRs
with different ribbon lengths $\ell$ in the $x$ direction (see \bluef{\ref{fig:Naming_conv_and_simul_cell}}).
The total energies of the relaxed structures are then fitted to a
polynomial function to obtain the optimized ribbon length $\ell_{0}$.
For ease of comparison between AGNR-$N$ and ZGNR-$N$, the optimized
lateral lattice parameter $a_{0}$ is calculated according to $a_{0}=\frac{\ell_{0}}{\sqrt{3}}$
and $a_{0}=\ell_{0}$ for AGNR-$N$ and ZGNR-$N$ respectively. From
\bluef{\ref{fig:Lattice_param}}, we find that while $a_{0}$ for
ZGNR-$N$ monotonically increases toward $a'_{0}=2.471$~\AA{}\ (the
optimized lattice parameter of graphene) with increasing $W$, $a_{0}$
of AGNR-$N$ monotonically decreases toward $a'_{0}$ with an observation
that AGNR-$N$ exhibits a three-family behavior for $a_{0}$, i.e.,
the convergence of $a_{0}$ is systematic when the AGNR-$N$ are grouped
according to $N~\mbox{modulo}~3$. We note that other three-family
behaviors for AGNR-$N$ have also been found for the electronic bandgap\citep{Son06v97},
edge energy\citep{Gan10v81,Wassmann2010}, and the LO/TO splitting\citep{Gillen09v80}.
The three-family behavior for $a_{0}$ of the AGNR-$N$ ribbons may
be understood using the concepts of aromaticity and resonance bond
theory.\citep{Randic2003} Wassmann \emph{et al.} argued that AGNRs
can be classified into three different families depending on the number
of equivalent Clar's structures that can be constructed for each AGNR-$N$.\citep{Wassmann2010}
Clar's structures must contain the maximum number of aromatic $\pi$-sextets
which can be accommodated by the structure. In \bluef{\ref{fig:clar_struc}},
we show examples of the equivalent Clar's structure that can be constructed
for AGNR-$N$ belonging to the three different families, and the bond
lengths for the optimized structures. For AGNR-$N$ where $N=3p$
and $p$ is an integer, only one Clar's structure can be constructed;
for $N=3p+1$, there are two equivalent Clar's structures, and for
$N=3p+2$, more than two equivalent Clar's structures can be constructed.
Since the C--C resonance bond is shorter than the C--C single bond,
the $N=3p$ structures will have longer bond lengths -- which results
in a larger $a_{0}$ -- as compared to the $N=3p+1$ and $N=3p+2$
structures. In contrast, for ZGNR-$N$ with unpaired spins at the
edges, more than two equivalent Clar's structure can be drawn for
any $N$.\citep{Wassmann2010} For width $W\sim50$~\AA{}, the value
of $a_{0}$ differs from that of the bulk graphene by less than $0.1$\%
($0.02$\%) for AGNR-$N$ (ZGNR-$N$).

\subsection{Thermal Conductance of Unstrained GNRs}

\begin{figure}
\begin{raggedleft} \includegraphics[clip,width=7.2cm]{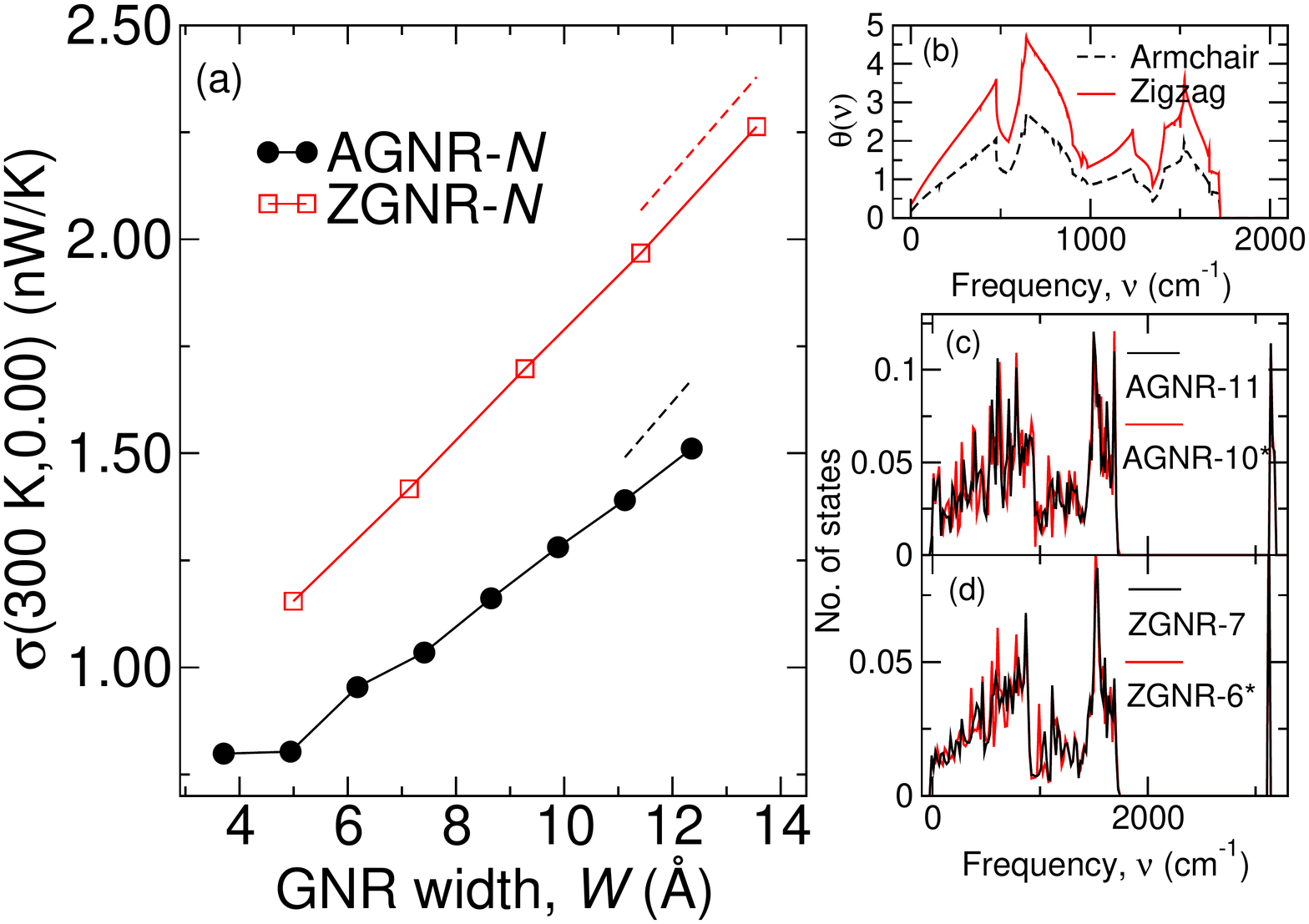} \caption{\textbf{(a)} Thermal conductance $\sigma(T=300~\mbox{K},\varepsilon=0.00)$
of AGNR-$N$ ($N=4$ to $11$) and ZGNR-$N$ ($N=3$ to $7$). The
slopes of the dash lines are deduced from the thermal conductance
of bulk graphene at 300~K in the armchair and zigzag directions.
\textbf{(b)} Average transmission function for bulk graphene along
the armchair and zigzag directions. \textbf{(c)} Phonon densities
of states (DOS) of AGNR-$11$ and AGNR-$10*$, which is the sum of
the DOS of AGNR-$10$ and the DOS of bulk graphene. \textbf{(d)} Phonon
DOS of ZGNR-$7$ and ZGNR-$6*$, which is the sum of the DOS of ZGNR-$6$
and the DOS of bulk graphene. \label{fig:conductance-of-unstrained-gnr-300K-and-bulk-graphene-theta} }

\end{raggedleft} 
\end{figure}

Using the optimized $\ell_{0}$ and atomic coordinates for GNR-$N$,
we perform the phonon dispersion calculation (see \bluef{\ref{fig:phonon}}
for typical results) and subsequently obtain the thermal conductance
by the counting method\citep{Markussen08v8}. \bluef{\ref{fig:conductance-of-unstrained-gnr-300K-and-bulk-graphene-theta}(a)}
shows the thermal conductance $\sigma(T,\varepsilon)$ at $T=300$~K
and $\varepsilon=0.00$ for GNRs as a function of $W$. We find that
ZGNR-$N$ have higher conductances as compared to AGNR-$N$ with comparable
$W$. This is due to the fact that the phonon dispersions of ZGNR
are more dispersive (a single phonon branch is more dispersive if it
covers a larger frequency range) compared to that of AGNRs,\citep{Tan11v11}
thus increasing the thermal conductance through a change in the transmission
function $\theta(\nu)$. Since the dispersiveness is related to the
gradient $d\omega/dk$, which is the phonon velocity,\citep{Rego98v81}
we may also say that ZGNRs have higher phonon velocities than AGNRs,
resulting in higher thermal conductance.

While bulk graphene is a fully $\pi$-resonant structure with equal
C--C bond lengths between the C atoms, the presence of edges in the
GNRs limits the extent of the $\pi$-resonance. Hence not all of the
C--C bond lengths are equivalent as explained in \bluef{\ref{fig:clar_struc}}.
Therefore, we expect the thermal conductance of GNRs to be different
from bulk graphene due to this edge effect. In \bluef{\ref{fig:conductance-of-unstrained-gnr-300K-and-bulk-graphene-theta}(b)},
we show the average transmission function $\theta(\nu)$ for the bulk
graphene in the armchair and zigzag directions. The $\sigma(300~\text{K},0.00)$
for bulk graphene in the armchair (zigzag) direction is $0.18$~nW/K
($0.32$~nW/K). As $W\rightarrow\infty$, the edge effect of GNRs
should converge to some finite value, and hence the conductance increase
from AGNR-$(N-1)$ to AGNR-$N$ should approach the conductance value
of bulk graphene in the armchair direction. The expected conductance
slopes for AGNRs and ZGNRs are shown in \bluef{\ref{fig:conductance-of-unstrained-gnr-300K-and-bulk-graphene-theta}(a)}.
\bluet{\ref{tab:cond-deviation}} shows that
for the largest $W$ investigated in this study --- AGNR-$11$
and ZGNR-$7$ --- the conductance slope for AGNR-11 is $34$\% lower
than that predicted for the bulk graphene, whereas the slope is only
$5$\% lower for ZGNR-$7$.
These discrepancies may be attributed to the strong edge effect on
the narrow width ribbons that we have studied here. Even though we
are unable to provide a quantitative measure of the edge effects,
we are able to provide qualitative evidence of the edge effect. In
\bluef{\ref{fig:conductance-of-unstrained-gnr-300K-and-bulk-graphene-theta}(c)},
we show the phonon density of states (DOS) for AGNR-$11$ and AGNR-$10*$,
which is sum of the DOS of AGNR-$10$ and the DOS of bulk graphene.
We find that the edge effect is still rather strong in the GNRs since
the DOS of AGNR-$11$ does not match well with AGNR-$10*$. Similarly
the DOS of ZGNR-$7$ does not match well with ZGNR-6{*}. This in turn
suggests why the conductance slopes do not agree for very narrow width
ribbons. We point out the fact that ZGNR-$N$ has a better agreement
in the conductance slope compared to AGNR-$N$ is consistent with
the observation that $a_{0}$ converges faster to $a'_{0}$ (the optimized
lattice parameter of graphene) for ZGRN-$N$ than for AGNR-$N$, as
shown in \bluef{\ref{fig:Lattice_param}}.

\begin{figure}
\begin{raggedleft} \includegraphics[clip,width=7.2cm]{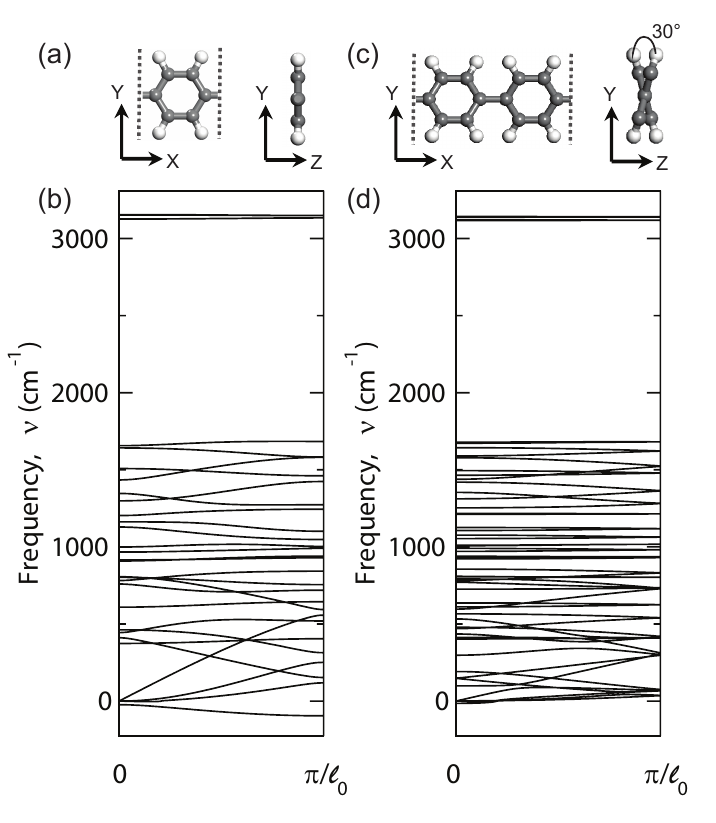} \caption{The primitive cell of AGNR-$3$ \textbf{(a)} shows a large negative
frequency of $\sim100$~$\text{cm}^{-1}$ in the phonon dispersion
relations at the zone boundary \textbf{(b)}. The relaxed structure
\textbf{(c)} of the enlarged primitive cell AGNR-$3$, where the torsion
angle between the 2 hexagon rings is $30^{\circ}$. \textbf{(d)} Phonon
dispersion relation of the enlarged primitive cell of AGNR-3 shows
no soft modes. \label{fig:AGNR-3_phonon} }

\end{raggedleft} 
\end{figure}

It is found that, except for AGNR-$3$, all unstrained ZGNR-$N$ and
AGNR-$N$ are stable when $l_{0}$ corresponds to $a_{0}'$ and $a_{0}'\sqrt{3}$,
respectively, with a small adjustment according to \bluef{\ref{fig:Lattice_param}}.
\bluef{\ref{fig:AGNR-3_phonon}(b)} shows the phonon dispersion of
AGNR-$3$ (more commonly known as polyphenylene) of a 10-atom primitive
cell that possesses soft modes extending from $\Gamma$ point to the
zone boundary with a large negative frequency of $\sim100$~$\text{cm}^{-1}$.
We create a new 20-atom supercell by combining two adjacent primitive
cells along the $x$ direction, which causes the soft mode of the
original supercell at the zone boundary to be folded to $\Gamma$
point of the Brillouin zone of the enlarged supercell. We perform
a phonon calculation for the 20-atom supercell and use the eigenvector
of the lowest frequency (i.e., the eigenmode with imaginary frequency)
to displace the atoms for a further atomic relaxation. We obtain a
relaxed structure shown in \bluef{\ref{fig:AGNR-3_phonon}(c)}, where
a torsion angle of $30^{\circ}$ between alternate hexagon rings is
observed. This is slightly smaller than the $40^{\circ}$ calculated
by Brocorens \emph{et al.} using the Hartree-Fock AM1 method for terphenyl.\citep{Brocorens1999}
The enlarged supercell for AGNR-3 is indeed stable according to the
phonon dispersion relation shown in \bluef{\ref{fig:AGNR-3_phonon}(d)},
where no soft modes are observed. This unit cell has a lower total
energy of $4$~meV/atom compared to that of the unstable, planar
AGNR-$3$.

\subsection{Effect of Tensile Strain on Thermal Conductance of GNRs}

\begin{figure}
\begin{raggedleft} \includegraphics[clip,width=6.2cm]{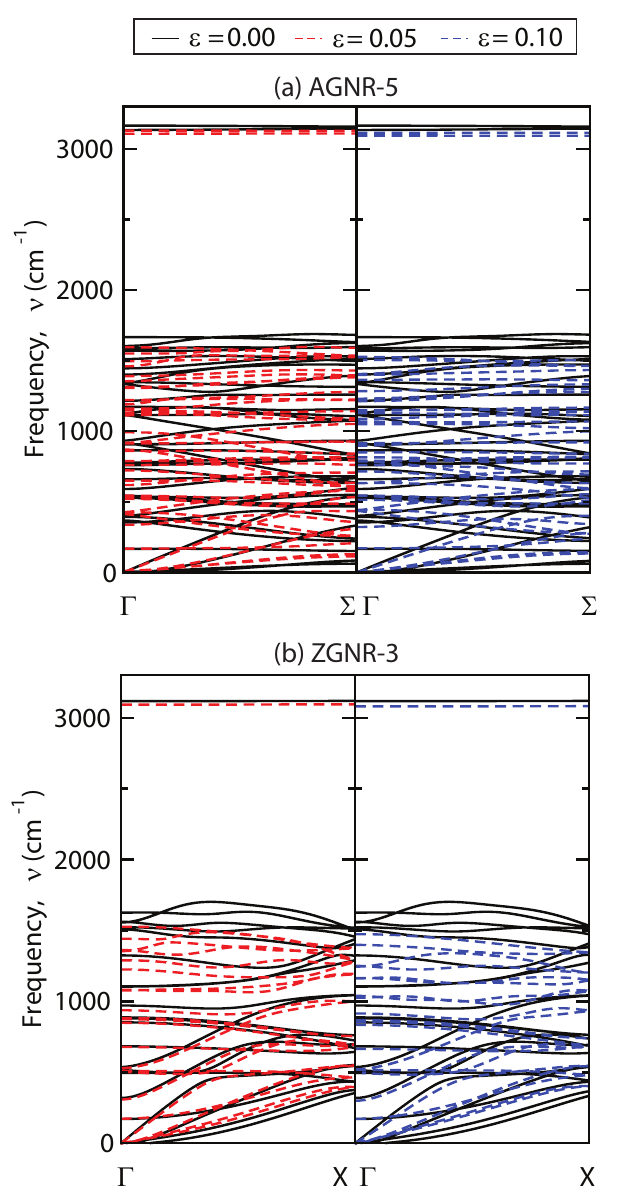}

\end{raggedleft}

\caption{Phonon dispersion relations for \textbf{(a)} AGNR-$5$ and \textbf{(b)}
ZGNR-$3$. Three different strain values of $\varepsilon=0.00$, $0.05$,
and $0.10$ are considered. \label{fig:phonon} }
\end{figure}

\begin{figure}
\begin{raggedleft} \includegraphics[clip,width=6.2cm]{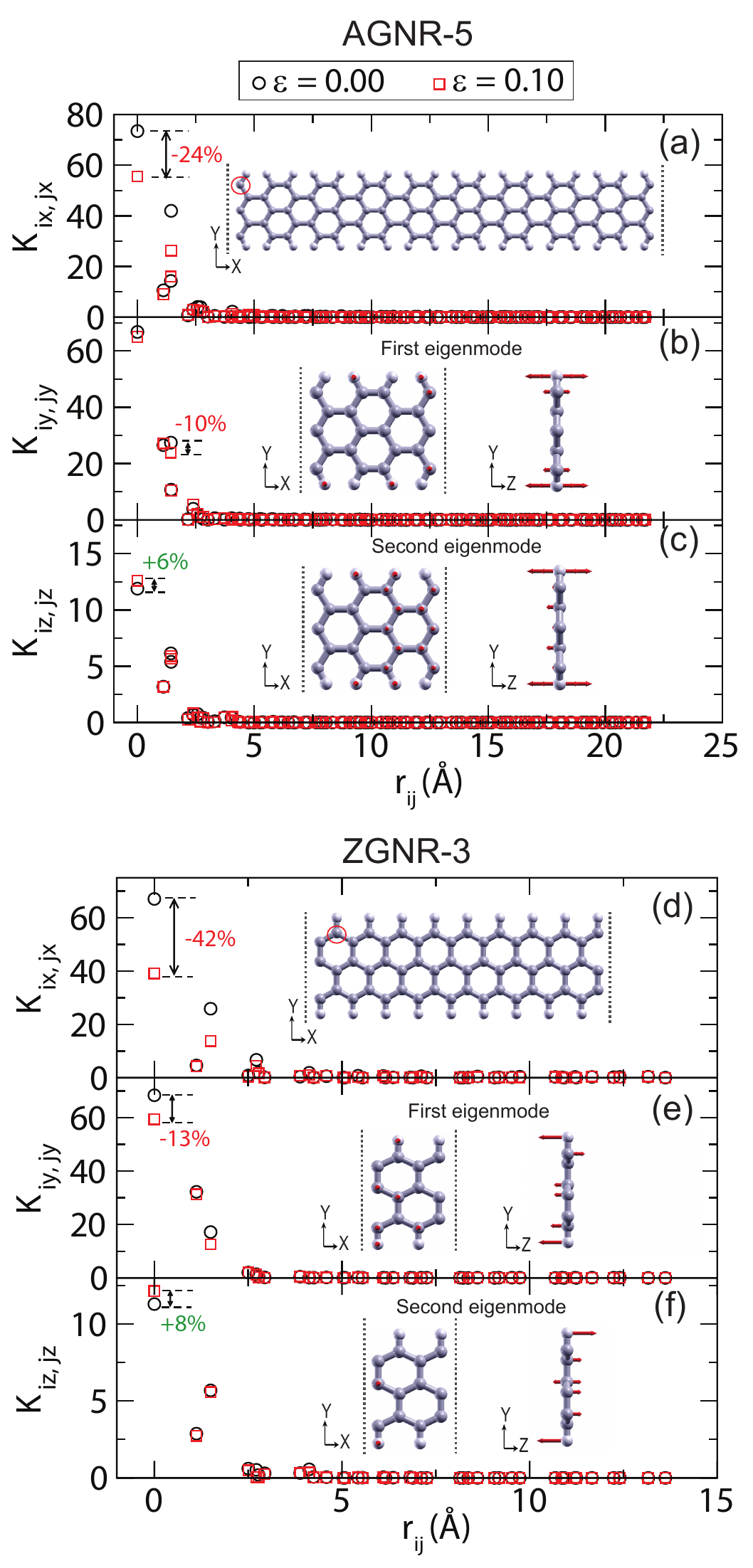} \caption{ The force constant $K_{i\alpha,\, j\alpha}$ (in units
eV/\AA{}$^{2}$) of AGNR-$5$ as a function of $r_{ij}=|{\bf r}_{i}-{\bf r}_{j}|$
for the $i$th ($j$th) atom located at ${\bf r}_{i}$ (${\bf r}_{j}$).
The $i$th atom (circled in red in the inset of \textbf{(a)}) is displaced
in \textbf{(a)} $\alpha=x$, \textbf{(b)} $\alpha=y$, and \textbf{(c)}
$\alpha=z$ direction. The strain values are $\varepsilon=0.00$ (black
circles) and $\varepsilon=0.10$ (red squares). The two lowest out-of-plane
(ZA) phonon eigenmodes are depicted in the insets of \textbf{(b)}
and \textbf{(c)}. The arrows (in red) on the atoms indicate the vibration
direction and amplitude of each atom. The results in \textbf{(d--f)}
are for ZNGR-$3$. \label{fig:fconst+eigenmode} }

\end{raggedleft} 
\end{figure}

Next we investigate the effect of tensile strain on the thermal conductance
of GNRs. Two different strain values $\epsilon=0.05$ and $0.10$
have been applied to the GNRs. Similar to the previous section, we
perform phonon calculations for AGNR-$N$ and ZGNR-$N$ before we
calculate the thermal conductance. \bluef{\ref{fig:phonon}} shows
the typical phonon dispersion relations for AGNR-$5$ and ZGNR-$3$.
We see that while the highest-frequency C--H stretching modes 
of $\sim3200$~\invcm\ are not affected by strain, 
all other high-frequency modes between
$\sim1000-1600$~\invcm\ are consistently reduced. On the other
hand, the two lowest-frequency out-of-plane acoustic (ZA) modes increase
in frequency as strain is applied. (In the case of graphene, Bonini
{\emph{et al.} observed
that the frequency of the out-of-plane mode increases as isotropic
tensile strain is applied.\citep{Bonini12v12}}) The changes to the
eigenmode frequencies may be understood from the changes to the force
constants as strain is applied, as shown in \bluef{\ref{fig:fconst+eigenmode}}
(to simplify analysis, we show only the $xx$, $yy$, and $zz$ diagonal
components). Tensile strain in AGNR-$5$ leads to substantial decrease
in the in-plane force constants along the longitudinal ($x$) and
transverse ($y$) directions due to the lengthened C--C bonds.\citep{Picu2003,Xu2009}
The largest decrease in the force constants for AGNR-$5$ is $-24$\%
and $-10$\% in the $xx$ and $yy$ directions, respectively. For ZGNR-$3$,
the largest decrease is $-42$\% and $-13$\% in the $xx$ and $yy$
components, respectively. However, the force constants in the out-of-plane
($z$) direction have either increased slightly or remained the same,
and this leads to the increase of the frequency of the out-of-plane
modes. The largest increase in the force constants in the $zz$ component
is $6$\% ($8\%$) for AGNR-$5$ (ZGNR-$3$).

\begin{figure}
\begin{raggedleft} \includegraphics[clip,width=7.2cm]{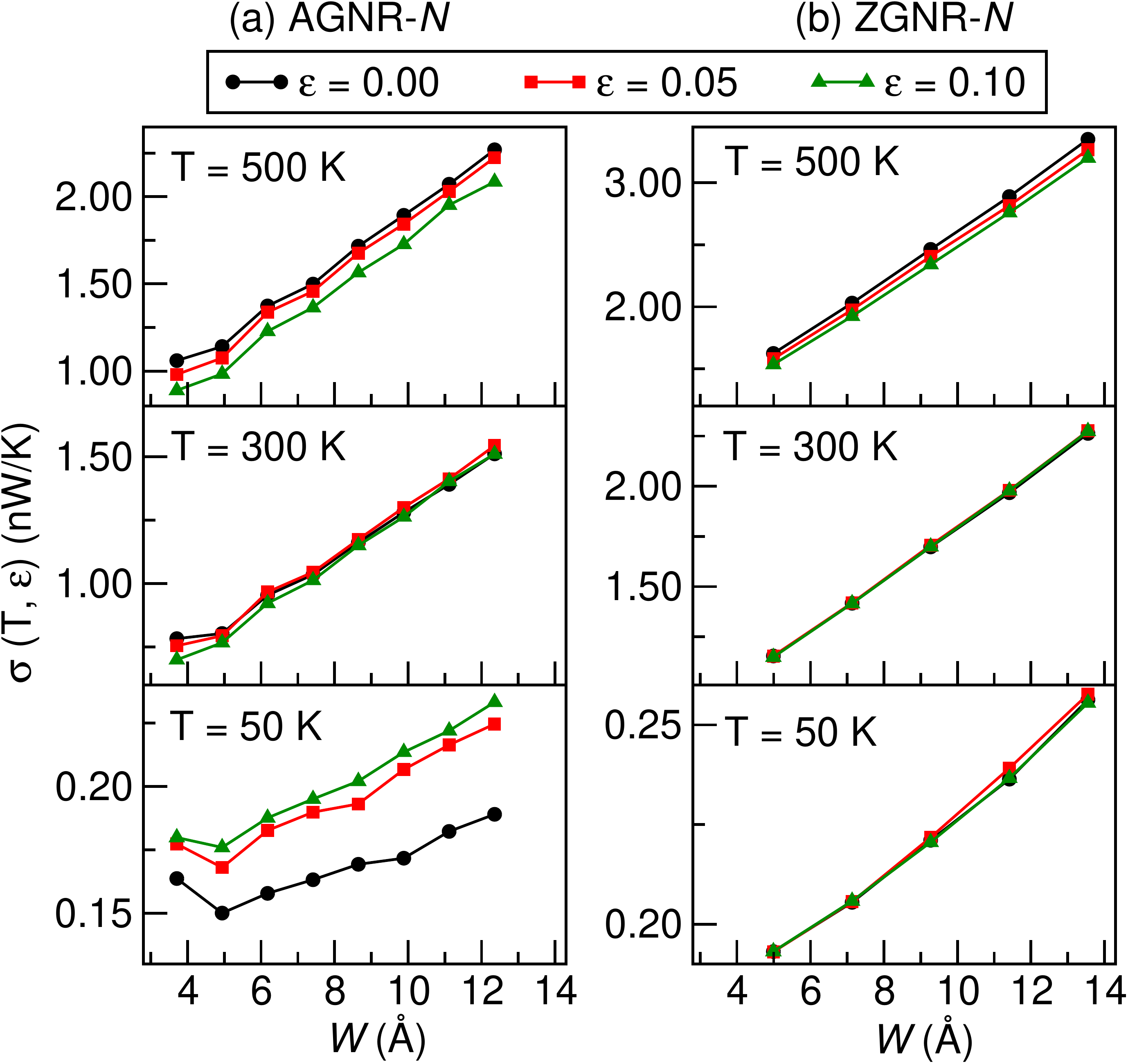} \caption{Conductance plots for \textbf{(a)} AGNR-$N$ ($N=4$ to $11$) and
\textbf{(b)} ZGNR-$N$ ($N=3$ to $7$) for $\epsilon=0.00,\;0.50\;\text{and}\;0.10$
and $T=50$, $300$ and $500$~K. \label{fig:width-gives-conductance-of-strained-AGNR-and-ZGNR-at-diff-T} }

\end{raggedleft} 
\end{figure}

\begin{figure}
\begin{raggedleft} \includegraphics[clip,width=7.2cm]{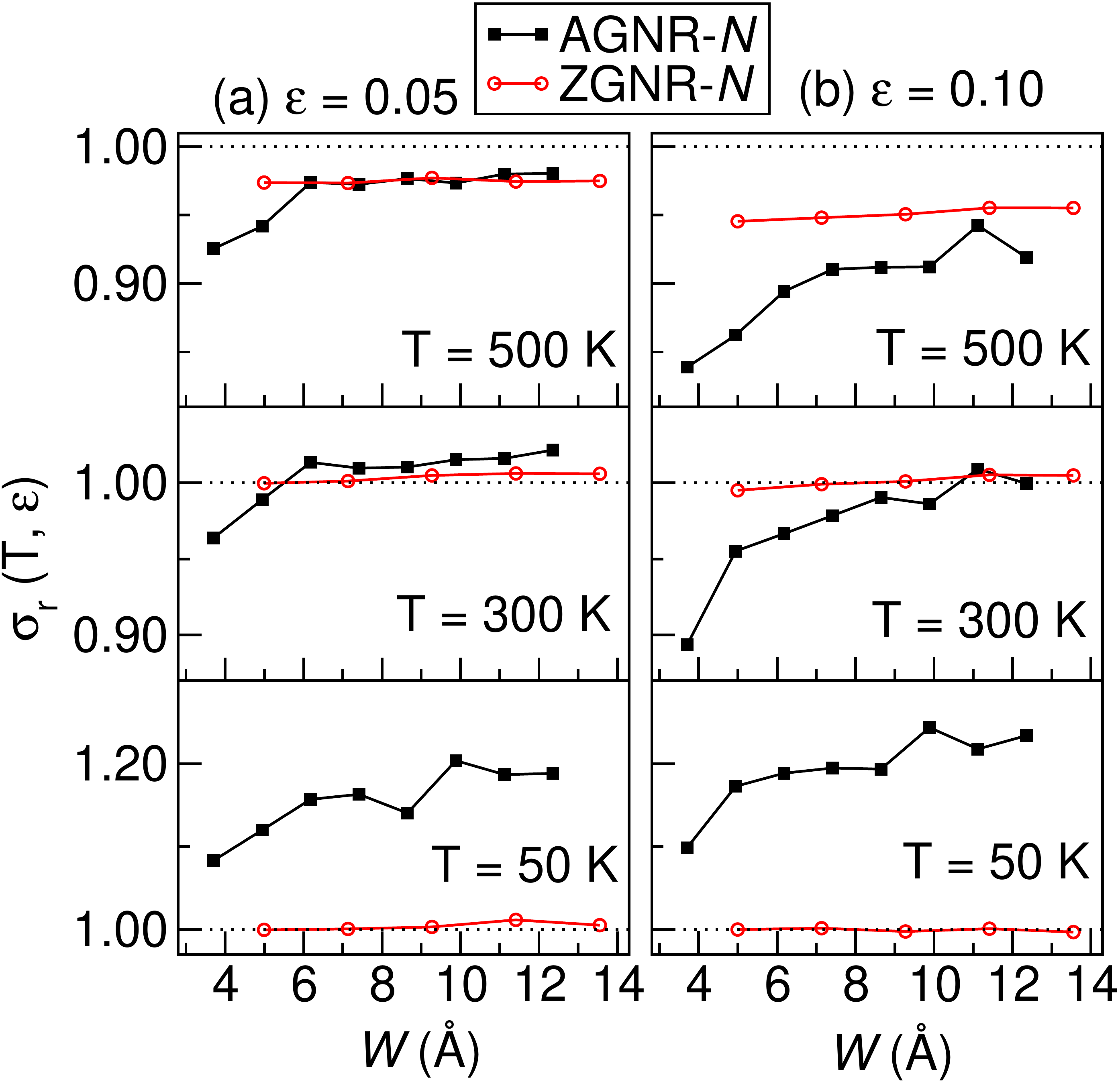} \caption{Relative conductance $\sigma_{\text{r}}(T,\varepsilon)$ of AGNR-$N$
($N=4$ to $11$) and ZGNR-$N$ ($N=3$ to $7$) of width $W$ for
\textbf{(a)} $\varepsilon=0.05$ and \textbf{(b)} $\varepsilon=0.10$
for at $T=50$, $300$ and $500$~K. \label{fig:relative-conductance-of-strained-AGNR-ZGNR-at-diff-T} }

\end{raggedleft} 
\end{figure}

The effect of strain on the thermal conductance is shown in \bluef{\ref{fig:width-gives-conductance-of-strained-AGNR-and-ZGNR-at-diff-T}}
for three representative temperatures $T=50$ (low temperature), $300$
(intermediate temperature) and $500$~K (high temperature). To directly
compare the conductance of a strained GNR and an unstrained GNR, we
define the relative conductance as $\sigma_{\text{r}}(T,\varepsilon)=\frac{\sigma(T,\varepsilon)}{\sigma(T,0)}$.
The results for $\sigma_{\text{r}}(T,\varepsilon)$ are shown in \bluef{\ref{fig:relative-conductance-of-strained-AGNR-ZGNR-at-diff-T}}.

We find that at a low temperature of $T=50$~K, the thermal conductance
is dominated by the low-frequency modes\citep{Tan11v11} since the
derivative $\frac{\partial n_{B}}{\partial T}$ diminishes rapidly
with increasing $\nu$. For AGNR-$N$, the thermal conductance substantially
increases when tensile strain is applied, where an increase of up
to $20$\% compared to the unstrained AGNR can be achieved. This increase
in conductance is due to an increase in the frequencies of the two
lowest-frequency acoustic phonon eigenmodes upon the application of
strain, as was discussed in \bluef{\ref{fig:phonon}(a)}.

However for ZGNR-$N$ at $T=50$~K, $\sigma_{\text{r}}(T,\varepsilon)$
with $\varepsilon=0.05\text{ and }0.10$ does not increase, but remains
fairly constant at 1, despite the fact that the lowest-frequency out-of-plane
ZA modes also increase in frequency after strain is applied. This
is because the two low-frequency out-of-plane ZA phonon eigenmodes
of unstrained ZGNR-$N$ are more dispersive than AGNR-$N$ (e.g.,
the frequency at the zone boundary is $370$~cm\textsuperscript{-1}
for ZGNR-$3$ versus $90$~cm\textsuperscript{-1} for AGNR-$5$,
as shown in \bluef{\ref{fig:phonon}}), even though the frequencies
of the lowest two ZA phonon branches increase with strain. Hence we
can understand that the increase of thermal conductance in AGNR-$N$
is due to the presence of new low-frequency modes that are easily
excited at a low temperature. For ZGNR-$N$, a low temperature only
excite the existing low-frequency modes because the new low-frequency
modes (from the effect of strain) occur at much higher frequencies
as compared to that of the AGNR-$N$, and therefore cannot be easily
excited thermally. Our results at low temperature differ from that
reported in Ref.~\citenum{Zhai2011} where essentially the same
increase in thermal conductance occurs for both AGNR-$N$ and ZGNR-$N$.
As the temperature increases, the higher-frequency modes start to
contribute to the thermal conductance in addition to the lower-frequency
ones. However, since the high-frequency modes are suppressed due to
the smaller force constants caused by the lengthening of the strained
bonds, there will be competition between the reduction in thermal
conductance due to the high-frequency modes, and the increase due
to low-frequency modes. For AGNR-$N$ at $T=300$ and $500$~K, the
former effect is stronger than the latter, especially at a larger
strain value, and this explains the significant reduction of the thermal
conductance as shown in \bluef{\ref{fig:width-gives-conductance-of-strained-AGNR-and-ZGNR-at-diff-T}(a)}
at progressively higher temperatures. However, for ZGNR-$N$, the
effects of high- and low-frequency almost cancel out each other, thus
leading to a near-constant behavior of thermal conductance.
From \bluet{\ref{tab:cond-deviation}}, we see that the edge effect 
is generally stronger for AGNRs than for ZGNRs even in the presence of strain.

%

\begin{figure}
\begin{raggedleft} \includegraphics[clip,width=6.7cm]{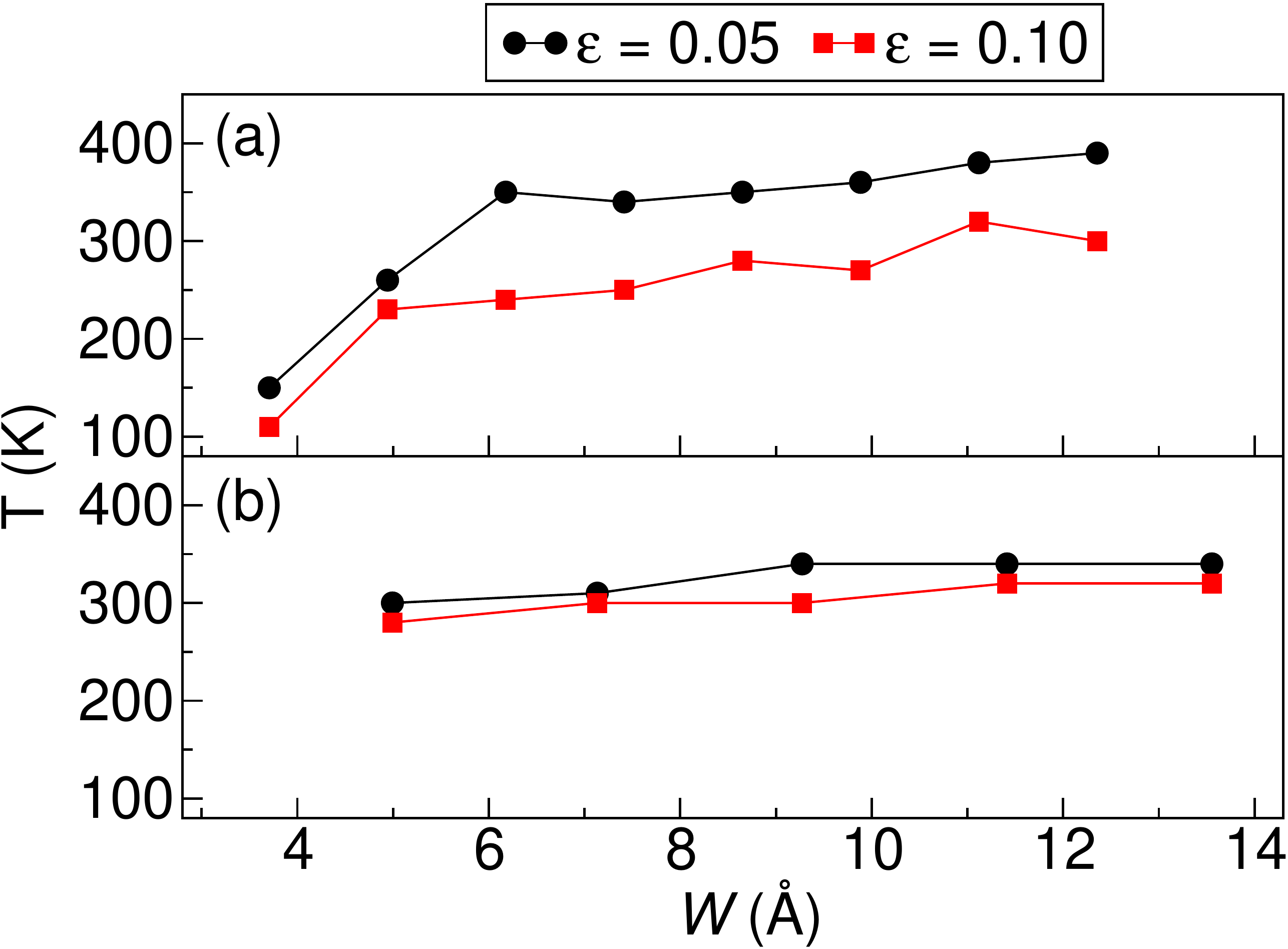} \caption{Cross-over temperatures (see text) for \textbf{(a)} AGNR-$N$ ($N=4$
to $11$) and \textbf{(b)} ZGNR-$N$ ($N=3$ to $7$) at $\varepsilon=0.05\text{ and }0.10$.
\label{fig:crossover-temperature-of-AGNR-and-ZGNR} }

\end{raggedleft} 
\end{figure}

Since the thermal conductance of the GNRs displays a nonmonotonic
behavior with temperature, it may be useful to discuss the `cross-over
temperature' of the GNRs, which is the temperature below (above) which
the thermal conductance of a strained GNR is more (less) than that
of the unstrained GNR. We show in \bluef{\ref{fig:crossover-temperature-of-AGNR-and-ZGNR}}
the cross-over temperatures of the strained GNRs, and note that they
are close to room temperature. This shows that we may control the
thermal conductance of strained GNRs within a reasonable range of
temperatures.

\begin{figure}
\begin{raggedleft} \includegraphics[clip,width=7.2cm]{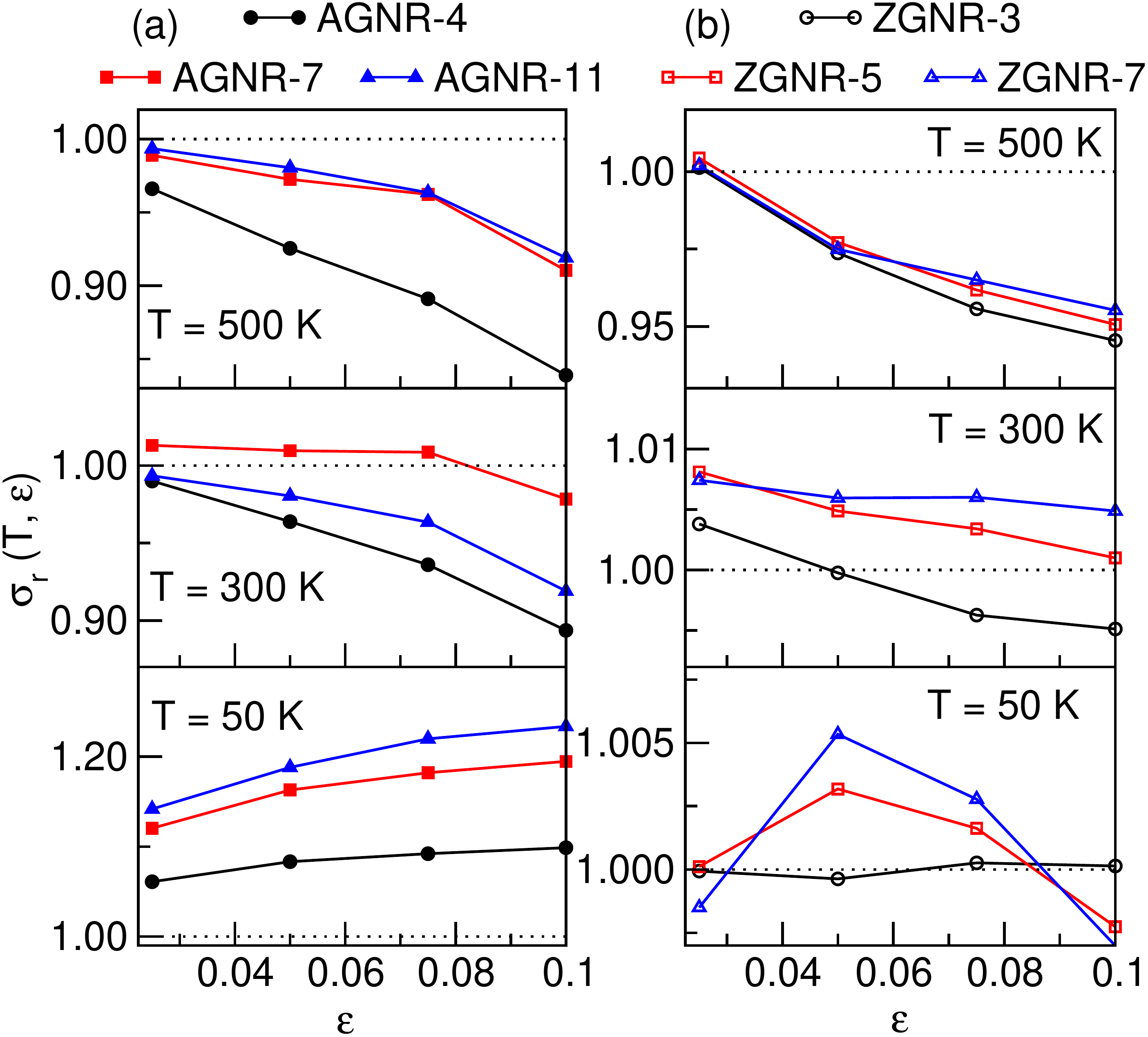} \caption{Relative conductance $\sigma_{\text{r}}(T,\varepsilon)$ versus strain
$\varepsilon$ for \textbf{(a)} AGNR-$N$ ($N=4$, $7$, and $11$)
and \textbf{(b)} ZGNR-$N$ ($N=3$, $5$, and $7$) at $T=50$, $300$
and $500$~K. \label{fig:relative-conductance-AGNR-and-ZGNR-different-strain-diff-T} }

\end{raggedleft} 
\end{figure}

Finally we also investigate the relative conductance of a few GNRs
at different strain values, the results of which are shown in \bluef{\ref{fig:relative-conductance-AGNR-and-ZGNR-different-strain-diff-T}}.
At low temperatures such as $T=50$~K, the thermal conductance of
AGNR-$N$ increases monotonically with increasing $\varepsilon$;
whereas it remains fairly constant for ZGNR-$N$. At higher temperatures
of $T=300$ and $500$~K, $\sigma_{\text{r}}(T,\varepsilon)$ for
both AGNR-$N$ and ZGNR-$N$ decrease with increasing $\varepsilon$,
with AGNR-$N$ showing a larger relative drop for the same amount
of $\varepsilon$. We find that the thermal conductance of AGNRs is
more sensitive to tensile strain than ZGNRs at $300$~K, which is
consistent with the conclusions of Wei \emph{et al.}.\citep{Wei11v22}
Zhai \emph{et al.}\citep{Zhai2011} predicted that with the application
of $\varepsilon=0.19$ on GNRs of wide widths ($\sim2.6$~nm), the
thermal conductance of strained GNRs should be higher than the corresponding
unstrained ones even at $400$~K. In the case of the narrow width
GNRs studied in this paper, we show in \bluef{\ref{fig:crossover-temperature-of-AGNR-and-ZGNR}}
that the cross-over temperatures do not exceed $400$~K. We attribute
the discrepancy between our results and those obtained by Zhai \emph{et
al.} due to their assumptions that the out-of-plane elastic constants
do not change with strain and that only nearest-neighbor force constants
are used. 
Recently, significant progress has been made toward synthesizing
GNRs of specific edge orientations.\citep{Cai2010}
Based on our results, we predict that the thermal conductance anisotropy
between AGNRs and ZGNRs suggests selective use of ZGNRs as fillers
in thermal interface materials because their thermal conductance is
higher and less adversely affected by strain as compared to AGNRs.
AGNRs on the other hand, might act as a suitable thermoelectric material
because of their inherent lower thermal conductance compared to ZGNRs,
and their thermal conductance can be further lowered with the application
of strain. However, the nonmonotonic variation of thermal conductance
with respect to strain for the AGNRs means that straining the AGNRs
to decrease their thermal conductance shall only be effective above
the cross-over temperature.

\section{Conclusion}

In conclusion, we have investigated the thermal conductance of AGNR-$N$
and ZGNR-$N$ as a function of tensile strain on the GNRs with accurate
density-functional theory calculations and the nonequilibrium Green's
function method. The lateral lattice constants of AGNR-$N$ are found
to follow a three-family behavior that depends on $N$. We found that
tensile strain decreases the force constants in the in-plane directions
of the GNRs, but slightly increases the force constants in the out-of-plane
direction, which increase the two lowest-frequency out-of-plane acoustic
modes. These modes play a decisive role in increasing the thermal
conductance of AGNR-$N$ at low temperatures, but not for ZGNR-$N$.
Hence the fact that the phonon dispersions of ZGNR-$N$ is more dispersive
than that of AGNR-$N$ of comparable widths results in two important
outcomes: (1) the unstrained ZGNR-$N$ has a higher thermal conductance
compared to the unstrained AGNR-$N$ of comparable widths, and (2)
that at low temperatures, the thermal conductance of the strained
ZGNR-$N$ is less sensitive toward the effect of strain than that
of AGNR-$N$. At high temperatures, the thermal conductance of the
strained ZGNR-$N$ and AGNR-$N$ decreases (relative to the unstrained
ones) with strain due to the fact that the frequency of the high-frequency
modes shifts down as a result of the
weakening of the interatomic force
constants in the in-plane direction. The use of the state-of-the-art
techniques such as density-functional theory and the nonequilibrium
Green's function method is able to reveal intricate quantum-mechanical
effects that may be hard or impossible to be captured with force-field
type interactions. Finally, it might be interesting to study the role
of out-of-plane modes in other strained single-layer systems such
BN, MoS$_{2}$, and WSe$_{2}$.

\section*{Acknowledgements}
The authors gratefully acknowledge useful discussions with
Jian-Sheng Wang, Zhen Wah Tan and Jinghua Lan. We acknowledge the
support of A{*}STAR Computational Resource Center (A{*}CRC) of Singapore.

\bibliographystyle{apsrev}

\begin{thebibliography}{10}

\bibitem{Dubi11v83}
Y.~Dubi and M.~DiVentra.
\newblock Colloquium: Heat flow and thermoelectricity in atomic and molecular
  junctions.
\newblock {\em Rev. Mod. Phys.}, 83:131, 2011.

\bibitem{Wang08v62}
Jian~{\mbox{-}S} Wang, J.~Wang, and J.~T. L\"u.
\newblock Quantum thermal transport in nanostructures.
\newblock {\em Eur. Phys. J. B}, 62(4):381, 2008.

\bibitem{Balandin11v10}
Alexander~A. Balandin.
\newblock Thermal properties of graphene and nanostructured carbon materials.
\newblock {\em Nat. Mater.}, 10(8):569, 2011.

\bibitem{Ghosh08v92}
S.~Ghosh, I.~Calizo, D.~Teweldebrhan, E.~P. Pokatilov, D.~L. Nika, A.~A.
  Balandin, W.~Bao, F.~Miao, and C.~N. Lau.
\newblock Extremely high thermal conductivity of graphene: Prospects for
  thermal management applications in nanoelectronic circuits.
\newblock {\em Appl. Phys. Lett.}, 92(15):151911, 2008.

\bibitem{Harman02v297}
T.~C Harman, P.~J Taylor, M.~P Walsh, and B.~E {LaForge}.
\newblock Quantum dot superlattice thermoelectric materials and devices.
\newblock {\em Science}, 297(5590):2229, 2002.

\bibitem{Balandin08v8}
A.~A. Balandin, S.~Ghosh, W.~Bao, I.~Calizo, D.~Teweldebrhan, F.~Miao, and
  C.~N. Lau.
\newblock Superior thermal conductivity of single-layer graphene.
\newblock {\em Nano Lett.}, 8:902, 2008.

\bibitem{Faugeras2010}
Clement Faugeras, Blaise Faugeras, Milan Orlita, M.~Potemski, Rahul~R. Nair,
  and A.~K. Geim.
\newblock Thermal conductivity of graphene in corbino membrane geometry.
\newblock {\em {ACS} Nano}, 4(4):1889, 2010.

\bibitem{Caiw2010}
Weiwei Cai, Arden~L. Moore, Yanwu Zhu, Xuesong Li, Shanshan Chen, Li~Shi, and
  Rodney~S. Ruoff.
\newblock Thermal transport in suspended and supported monolayer graphene grown
  by chemical vapor deposition.
\newblock {\em Nano Lett.}, 10(5):1645, 2010.

\bibitem{Yu2007}
Aiping Yu, Palanisamy Ramesh, Mikhail~E. Itkis, Elena Bekyarova, and Robert~C.
  Haddon.
\newblock Graphite {NanoplateletsEpoxy} composite thermal interface materials.
\newblock {\em J. Phys. Chem. C}, 111(21):7565, 2007.

\bibitem{Fang2010}
Ming Fang, Kaigang Wang, Hongbin Lu, Yuliang Yang, and Steven Nutt.
\newblock Single-layer graphene nanosheets with controlled grafting of polymer
  chains.
\newblock {\em J. Mater. Chem.}, 20(10):1982, 2010.

\bibitem{Bolotin2008}
K~I Bolotin, K~J Sikes, J~Hone, H~L Stormer, and P~Kim.
\newblock {Temperature-Dependent} transport in suspended graphene.
\newblock {\em Phys. Rev. Lett.}, 101(9):096802, 2008.

\bibitem{Dragoman2007}
D.~Dragoman and M.~Dragoman.
\newblock Giant thermoelectric effect in graphene.
\newblock {\em Appl. Phys. Lett.}, 91:203116, 2007.

\bibitem{Wei2009}
Peng Wei, Wenzhong Bao, Yong Pu, Chun~Ning Lau, and Jing Shi.
\newblock Anomalous thermoelectric transport of dirac particles in graphene.
\newblock {\em Phys. Rev. Lett.}, 102(16):166808, 2009.

\bibitem{Hsu2004}
Kuei~Fang Hsu, Sim Loo, Fu~Guo, Wei Chen, Jeffrey~S Dyck, Ctirad Uher, Tim
  Hogan, E.~K Polychroniadis, and Mercouri~G Kanatzidis.
\newblock Cubic {AgPbmSbTe2+m:} bulk thermoelectric materials with high figure
  of merit.
\newblock {\em Science}, 303(5659):818, 2004.

\bibitem{Klemens2000a}
P.~G. Klemens.
\newblock Theory of the {A}-plane thermal conductivity of graphite.
\newblock {\em J. Wide Bandgap Mater.}, 7(4):332, 2000.

\bibitem{Savin2010}
Alexander~V. Savin, Yuri~S. Kivshar, and Bambi Hu.
\newblock Suppression of thermal conductivity in graphene nanoribbons with
  rough edges.
\newblock {\em Phys. Rev. B}, 82(19):195422, 2010.

\bibitem{Hu2010}
Jiuning Hu, Stephen Schiffli, Ajit Vallabhaneni, Xiulin Ruan, and Yong~P Chen.
\newblock Tuning the thermal conductivity of graphene nanoribbons by edge
  passivation and isotope engineering: A molecular dynamics study.
\newblock {\em Appl. Phys. Lett.}, 97(13):133107, 2010.

\bibitem{Haskins11v5}
J.~Haskins, A.~Kinaci, Cem Sevik, H.~Sevincli, G.~Cuniberti, and T.~Cagin.
\newblock Control of thermal and electronic transport in defect-engineered
  graphene nanoribbons.
\newblock {\em ACS Nano}, 5:3779, 2011.

\bibitem{Gunst2011}
Tue Gunst, Troels Markussen, A.~{\mbox{-P}}. Jauho, and Mads Brandbyge.
\newblock Thermoelectric properties of finite graphene antidot lattices.
\newblock {\em Phys. Rev. B}, 84(15):155449, 2011.

\bibitem{Ghosh09v11}
S.~Ghosh, D.~L. Nika, E.~P. Pokatilov, and A.~A. Balandin.
\newblock Heat conduction in graphene: experimental study and theoretical
  interpretation.
\newblock {\em New J. Phys.}, 11:095012, 2009.

\bibitem{Nika09v79}
D.~L. Nika, E.~P. Pokatilov, A.~S. Askerov, and A.~A. Balandin.
\newblock Phonon thermal conduction in graphene: Role of umklapp and edge
  roughness scattering.
\newblock {\em Phys. Rev. B}, 79:155413, 2009.

\bibitem{Lan09v79}
J.~Lan, J.~{\mbox{-S}}. Wang, C.~K. Gan, and S.~K. Chin.
\newblock Edge effects on quantum thermal transport in graphene nanoribbons:
  Tight-binding calculations.
\newblock {\em Phys. Rev. B}, 79:115401, 2009.

\bibitem{Xie11v23}
Zhong~Xiang Xie, Ke~Qiu Chen, and Wen~Hui Duan.
\newblock Thermal transport by phonons in zigzag graphene nanoribbons with
  structural defects.
\newblock {\em J. Phys.: Condens. Matter}, 23:315302, 2011.

\bibitem{Guo09v95}
Z.~X. Guo, D.~E. Zhang, and X.~G. Gong.
\newblock Thermal conductivity of graphene nanoribbons.
\newblock {\em Appl. Phys. Lett.}, 95:163103, 2009.

\bibitem{Wei11v22}
N.~Wei, L.~Xu, H.~Q. Wang, and J.~C. Zheng.
\newblock Strain engineering of thermal conductivity in graphene sheets and
  nanoribbons: a demonstration of magic flexibility.
\newblock {\em Nanotechnology}, 22:105705, 2011.

\bibitem{Hu09v9}
J.~Hu, C.~Ruan, and Y.~P. Chen.
\newblock Thermal conductivity and thermal rectification in graphene
  nanoribbons: A molecular dynamics study.
\newblock {\em Nano Lett.}, 9:2730, 2009.

\bibitem{Munoz10v10}
Enrique Mu\~{n}oz, Jianxin Lu, and Boris~I. Yakobson.
\newblock Ballistic thermal conductance of graphene ribbons.
\newblock {\em Nano Letters}, 10(5):1652--1656, 2010.

\bibitem{Tan11v11}
Z.~W. Tan, J.~{\mbox{-S}}. Wang, and C.~K. Gan.
\newblock {First-Principles} study of heat transport properties of graphene
  nanoribbons.
\newblock {\em Nano Lett.}, 11(1):214, 2011.

\bibitem{Sun08v129}
L.~Sun, Q.~X. Li, H.~Ren, H.~B. Su, Q.~W. Shi, and J.~L. Yang.
\newblock Strain effect on electronic structures of graphene nanoribbons: A
  first-principles study.
\newblock {\em J. Chem. Phys.}, 129:074704, 2008.

\bibitem{Kim09v457}
Keun~Soo Kim, Yue Zhao, Houk Jang, Sang~Yoon Lee, Jong~Min Kim, Kwang~S. Kim,
  Jong-Hyun Ahn, Philip Kim, Jae-Young Choi, and Byung~Hee Hong.
\newblock Large-scale pattern growth of graphene films for stretchable
  transparent electrodes.
\newblock {\em Nature}, 457:706, 2009.

\bibitem{Huang09v106}
M.~Y. Huang, H.~G. Yen, C.~Y. Chen, D.~H. Song, Tony~F. Heinz, and J.~Hone.
\newblock Phonon softening and crystallographic orientation of strained
  graphene studied by raman spectroscopy.
\newblock {\em Proc. Nat. Acad. Sci. USA}, 106:7304, 2009.

\bibitem{Li10v81}
Xiaobo Li, Kurt Maute, Martin~L. Dunn, and Ronggui Yang.
\newblock Strain effects on the thermal conductivity of nanostructures.
\newblock {\em Phys. Rev. B}, 81(24):245318, 2010.

\bibitem{Gunawardana12v85}
K.~G. S.~H. Gunawardana, Kieran Mullen, Jiuning Hu, Yong~P. Chen, and Xiulin
  Ruan.
\newblock Tunable thermal transport and thermal rectification in strained
  graphene nanoribbons.
\newblock {\em Physical Review B}, 85(24):245417, 2012.

\bibitem{Sevincli10v81}
H.~Sevinçli and G.~Cuniberti.
\newblock Enhanced thermoelectric figure of merit in edge-disordered zigzag
  graphene nanoribbons.
\newblock {\em Phys. Rev. B}, 81(11):113401, March 2010.

\bibitem{Sevincli10v82}
Wu~Li, H.~Sevinçli, G.~Cuniberti, and Stephen Roche.
\newblock Phonon transport in large scale carbon-based disordered materials:
  Implementation of an efficient order-n and real-space {K}ubo methodology.
\newblock {\em Phys. Rev. B}, 82:041410, 2010.

\bibitem{Zhai2011}
X.~Zhai and G.~Jin.
\newblock {Stretching-enhanced} ballistic thermal conductance in graphene
  nanoribbons.
\newblock {\em Europhys. Lett.}, 96:16002, 2011.

\bibitem{Wang06v74}
Jian~{\mbox{-}S} Wang, Jian Wang, and Nan Zeng.
\newblock Nonequilibrium {Green's} function approach to mesoscopic thermal
  transport.
\newblock {\em Phys. Rev. B}, 74(3):033408, 2006.

\bibitem{Wang07v75}
Jian~{\mbox{-}S} Wang, Nan Zeng, Jian Wang, and Chee~Kwan Gan.
\newblock Nonequilibrium {Green's} function method for thermal transport in
  junctions.
\newblock {\em Phys. Rev. E}, 75(6):061128, 2007.

\bibitem{Mingo06v74}
N.~Mingo.
\newblock Anharmoic phonon flow through molecular-sized junctions.
\newblock {\em Phys. Rev. B}, 74:125402, 2006.

\bibitem{Brandbyge02v65}
Mads Brandbyge, Jose-Luis Mozos, Pablo Ordejon, Jeremy Taylor, and Kurt
  Stokbro.
\newblock Density-functional method for nonequilibrium electron transport.
\newblock {\em Phys. Rev. B}, 65:165401, 2002.

\bibitem{Huang10v108}
Zhen Huang, Timothy~S. Fisher, and Jayathi~Y. Murthy.
\newblock Simulation of phonon transmission through graphene and graphene
  nanoribbons with a green's function method.
\newblock {\em Journal of Applied Physics}, 108(9):094319, November 2010.

\bibitem{Gan10v81}
C.~K. Gan and D.~J. Srolovitz.
\newblock First-principles study of graphene edge properties and flake shapes.
\newblock {\em Phys. Rev. B}, 81(12):125445, 2010.

\bibitem{Bao09v4}
Wenzhong Bao, Feng Miao, Zhen Chen, Hang Zhang, Wanyoung Jang, Chris Dames, and
  Chun~Ning Lau.
\newblock {Controlled ripple texturing of suspended graphene and ultrathin
  graphite membranes}.
\newblock {\em {Nature Nanotechnology}}, {4}({9}):{562--566}, {2009}.

\bibitem{Kumar10v82}
Sandeep Kumar, K.~P. S.~S. Hembram, and Umesh~V. Waghmare.
\newblock Intrinsic buckling strength of graphene: First-principles density
  functional theory calculations.
\newblock {\em Phys. Rev. B}, 82:115411, 2010.

\bibitem{Markussen08v8}
T.~Markussen, A.~{\mbox{-P}}. Jauho, and Mads Brandbyge.
\newblock Heat conductance is strongly anisotropic for pristine silicon
  nanowires.
\newblock {\em Nano Lett.}, 8:3771, 2008.

\bibitem{Rego98v81}
L.~G.~C. Rego and G.~Kirczenow.
\newblock Quantized thermal conductance of dielectric quantum wires.
\newblock {\em Phys. Rev. Lett.}, 81:232, 1998.

\bibitem{Yamamoto04v92}
T.~Yamamoto, S.~Watanabe, and K.~Watanabe.
\newblock Universal features of quantized thermal conductance of carbon
  nanotubes.
\newblock {\em Phys. Rev. Lett.}, 92:075502, 2004.

\bibitem{Soler2002}
J.~Soler, E.~Artacho, J.~D. Gale, A.~Garc\'{i}a, J.~Junquera, P.~Ordej\'{o}n,
  and D.~S\'{a}nchez-Portal.
\newblock The {SIESTA} method for ab initio {order-N} materials simulation.
\newblock {\em J. Phys.: Condens. Matter}, 14:2745, 2002.

\bibitem{Son06v97}
Y.~W. Son, M.~L. Cohen, and Steven~G. Louie.
\newblock Energy gaps in graphene nanoribbons.
\newblock {\em Phys. Rev. Lett.}, 97:216803, 2006.

\bibitem{Kresse1995}
G.~Kresse, J.~Furthmüller, and J.~Hafner.
\newblock Ab initio force constant approach to phonon dispersion relations of
  diamond and graphite.
\newblock {\em Europhys. Lett.}, 32:729, 1995.

\bibitem{Gan06v73}
C.~K. Gan, Y.~P. Feng, and D.~J. Srolovitz.
\newblock First-principles calculation of the thermodynamics of {In(x)Ga(1-x)N}
  alloys: Effect of lattice vibrations.
\newblock {\em Phys. Rev. B}, 73:235214, 2006.

\bibitem{Zhao11v84}
Y.~Y. Zhao, K.~T.~E. Chua, C.~K. Gan, J.~Zhang, B.~Peng, Z.~P. Peng, and Q.~H.
  Xiong.
\newblock Phonons in {Bi2S3} nanostructures: Raman scattering and
  first-principles studies.
\newblock {\em Phys. Rev. B}, 84:205330, 2011.

\bibitem{Wassmann2010}
Tobias Wassmann, Ari~P. Seitsonen, A.~Marco Saitta, Michele Lazzeri, and
  Francesco Mauri.
\newblock Clar's theory, {Pi-Electron} distribution, and geometry of graphene
  nanoribbons.
\newblock {\em J. Am. Chem. Soc.}, 132(10):3440, 2010.

\bibitem{Gillen09v80}
Roland Gillen, Marcel Mohr, Christian Thomsen, and Janina Maultzsch.
\newblock Vibrational properties of graphene nanoribbons by first-principles
  calculations.
\newblock {\em Phys. Rev. B}, 80(15):155418, 2009.

\bibitem{Randic2003}
M.~Randic.
\newblock Aromaticity of polycyclic conjugated hydrocarbons.
\newblock {\em Chem. Rev.}, 103(9):3449, 2003.

\bibitem{Brocorens1999}
P.~Brocorens, E.~Zojer, J.~Cornil, Z.~Shuai, G.~Leising, K.~Müllen, and {JL}
  Brédas.
\newblock Theoretical characterization of phenylene-based oligomers, polymers,
  and dendrimers.
\newblock {\em Synth. Met.}, 100(1):141, 1999.

\bibitem{Bonini12v12}
Nicola Bonini, Jivtesh Garg, and Nicola Marzari.
\newblock Acoustic phonon lifetimes and thermal transport in free-standing and
  strained graphene.
\newblock {\em Nano Letters}, 12(6):2673--2678, June 2012.

\bibitem{Picu2003}
R.~C. Picu, T.~{Borca-Tasciuc}, and M.~C. Pavel.
\newblock Strain and size effects on heat transport in nanostructures.
\newblock {\em J. Appl. Phys.}, 93(6):3535, 2003.

\bibitem{Xu2009}
Zhiping Xu and Markus~J Buehler.
\newblock Strain controlled thermomutability of single-walled carbon nanotubes.
\newblock {\em Nanotechnology}, 20(18):185701, May 2009.

\bibitem{Cai2010}
J.~Cai, P.~Ruffieux, R.~Jaafar, M.~Bieri, T.~Braun, S.~Blankenburg, M.~Muoth,
  A.~P Seitsonen, M.~Saleh, and X.~Feng.
\newblock Atomically precise bottom-up fabrication of graphene nanoribbons.
\newblock {\em Nature}, 466(7305):470, 2010.

\end{thebibliography}

\end{document}